\documentclass[letterpaper]{jpconf}
\usepackage{graphicx}
\usepackage{graphicx}
\usepackage{dcolumn}
\begin{document}

\title{Search for a Light Standard Model Higgs Boson Produced in Association with a Photon in Vector Boson Fusion}

\author{D. Asner$^1$, M. Cunningham$^1$, S. Dejong$^1$, K. Randrianarivony$^1$, C. Santamarina R\'{\i}os$^2$,  M. Schram$^2$} 

\address{$^1$ Carleton University, Ottawa, ON, Canada} 
\address{$^2$ McGill University, Montreal, QC, Canada}
\ead{krandria@physics.carleton.ca}


\begin{abstract}
We present the prospects for the discovery of a Standard Model Higgs boson
decaying to $b\bar{b}$ pair through the Vector Boson Fusion mechanism
in association with a central photon at the Large Hadron Collider (LHC). The photon provides a clean trigger and several advantages despite the fact that the cross section decreases  by approximately a factor of $\alpha_{em}$. After applying various optimized selection criteria, we obtain a
significance of $\sim$1.9 using ALPGEN interfaced with PYTHIA. The expected $b$-jet efficiency and fake rates and other detector acceptance cuts are taken into account to obtain a more realistic result.

\end{abstract}

\section{Introduction}

The primary purpose of the LHC is to study the electroweak symmetry breaking mechanism. It is described in the Standard Model (SM) by the Higgs mechanism
which includes a scalar Higgs Boson. The largest branching ratio for a low mass Higgs is
$h\rightarrow b\bar{b}$. This process could potentially help in determining spin, CP, gauge coupling, and Yukawa coupling of the Higgs candidates. Gluon-fusion, which is the largest production mechanism, is inaccessible due to its large background. The second largest production mode, Vector-Boson-Fusion (VBF), will be challenging as there is presently no trigger strategy. A parton level study by E. Gabirelli {\em et al.}~\cite{Mele07} suggests a statistical signifcance $S/\sqrt{B}$ around $3$ using an associated central photon to the VBF production. Despite the fact that  requiring an associated photon decreases the cross section by approximately a factor of
$\alpha_{em}$, it provides a clean and efficient trigger. It also suppresses the gluonic background as gluons do not radiate a photon and suppresses the $hZZ$ signal contribution allowing for possible measurements of the $hbb$ and $hWW$ couplings. We investigated the sensitivity by studying the effects originating from hadronization/showering, the Underlying Event (UE) model, and jet performance on the discovery potential. 

\section{Signal and Background}
There are two stages to create the signal and background samples. First, we generated the 4-vector samples using ALPGEN~\cite{Manga02}. Then, PYTHIA~\cite{pythia} was used to perform the hadronization, showering, UE,  and decays. HDECAY \cite{Djouadi97} evaluates the signal branching fraction. The same final state partons $b\bar{b}$+2jets+$\gamma$ is our primary background. Due to a possible contamination of the high-side of the $Z$-boson mass peak, we considered $Z$ with multiple jets and a photon. We also considered QCD multijets with an associated photon. The cross sections for each sample are shown in Table~\ref{xsec}.

\begin{table}[htpb]
\begin{center}
\begin{tabular}{lcr}

\begin{tabular}{c|c}
\hline \hline
$\sigma(h\gamma jj)$ & 69.74 $(fb)$ \\
BR($h\rightarrow b\bar{b}$) & 0.731 \\
$\sigma \times BR$ & 50.98 $(fb)$\\ \hline \hline
\end{tabular}

&
\hspace{2cm}
&

{\scriptsize
\begin{tabular}{c|c|c|c|c}
\hline
& Parton $\#$ & $\sigma (pb)$ & MLM $\epsilon$ (\%)  & $\sigma ' (pb) $\\ \hline 
$b\bar{b}$+$n$parton+$\gamma$ & 1 & 1088 & 38  & 413 \\
& 2 &  658 & 36 & 235 \\
\hline
$n$parton+$\gamma$ & 3 & 45789 & 20  & 9158 \\
& 4 & 17595 & 18 & 3167 \\
\hline
$Z$+$n$parton+$\gamma$ & 1 & 27 & 41  & 11 \\
& 2 & 18 & 44 & 8 \\
\hline
\end{tabular}
}
\\
\end{tabular}
\end{center}
\caption{Left: Signal cross-section and branching fraction for a 115 GeV/c$^2$ Higgs mass. Right: Background cross-sections, matching efficiencies, and the effective cross-sections.}
\label{xsec}
\end{table}

\section{Jet Reconstruction}

We use particles after hadronization and showering as the jet components to study its effects on the potential to observe our signal. We have studied the following algorithms to determine which one is preferred: the SISCone with 0.4 and 0.7 cone radius and AntiKt algorithms with 0.4 and 0.6 aggregation distances~\cite{sjet}. We based our study on five quantities: efficiency of individual jets, the jet pair efficiency as a function of the initial parton's distance, the transverse energy linearity, the transverse energy resolution and the jet multiplicity after reconstruction. The AntiKt algorithm with 0.6 aggregation parameter has the best overall performance and was used for the rest of the analysis.

\section{Systematic Uncertainties}

We studied the effects of three sources of uncertainty: choice of factorization and renormalization, Monte Carlo tunes and input Parton Density Functions (PDFs). The choice of the factorization and renormalization gave an uncertainty of $5\%$ for the signal and ranges from $10\%$ to $60\%$ for the backgrounds. The PDFs effects were studied by using the full group 40 CTEQ6M and gave an uncertainty of the order of $4\%$ both for signal and background. As for the systematic study originating from showering, hadronization and the UE models, we considered variations from the nominal tune (Perugia 0) and two other tunes (Perugia `soft' and `hard'). It yielded an uncertainty of $15.6\%$ and $26.5\%$ for signal and background, respectively, however, the impact on the overall significance was approximately $4\%$.
     
\section{Results}

We selected events based on the following selection criteria: a photon with $p_{\rm{T}}>30$ GeV/c and four or more jets in the event. A VBF likelihood tagger has been developped based on the geometrical properties of the VBF quarks to select VBF jets. We also applied kinematic selections such as their transverse momenta, invariant mass, and the geometrical separation between the two VBF jets. The $b$-jets are identified based on the truth information. We assumed $60\%$ $b$-tagging efficiency which corresponds to a 10 and 200 rejection factors for $c$-jets and light jets (jets originating from $u, d, s$ quarks and gluons), respectively. After we selected the VBF and $b$-jets candidates, we applied a veto on events with any remaining central jets, $|\eta| < 2$ and $p_{\rm{T}}>25$ GeV/c. Then, we considered a mass window range near the nominal signal Higss mass. The expected number of events after each selection criteria are shown in Table~\ref{tresults}. We obtained a sensitivity $S/\sqrt{B}\sim1.9$ with 100 fb$^{-1}$ accumulated data for a 115 GeV/c$^{2}$ SM Higgs mass. In addition, we show on Table~\ref{t:vbfph_sig_mass} the expected number of signal and background events and their respective significances for 115, 125 and 135~GeV/$c^2$ Higgs mass. Note that our result is comparable with the  
anticipated results from $W/Z h(\rightarrow b\bar{b})$~\cite{WHbb} with ~50fb$^{-1}$ of LHC data. As other $h\rightarrow b\bar{b}$ modes do not give overwhelming significance, we should include VBF $h(\rightarrow b\bar{b})+\gamma$ to enable the observation of a light SM Higgs. Moreover, this channel will give us the possibility of measuring $hbb$ and $hWW$ couplings.

\begin{table*}[htdp] 
\begin{center} 
{\scriptsize
\begin{tabular}{|l|rr|@{}rr@{}|@{}rr@{}|@{}rr@{}|@{}rr@{}|@{}rr@{}|@{}rr@{}|}\hline 
Cut \# &\multicolumn{2}{c|}{$m_h$ 115 GeV/$c^2$} & \multicolumn{2}{c|}{$b\bar{b}$+2jets+$\gamma$} & \multicolumn{2}{c|}{$b\bar{b}$+1jet+$\gamma$} & \multicolumn{2}{c|}{4jets+$\gamma$} & \multicolumn{2}{c|}{3jets+$\gamma$} & \multicolumn{2}{c|}{z+2jets+$\gamma$} & \multicolumn{2}{c|}{z+1jet+$\gamma$} \\ \hline 
None & \multicolumn{2}{l|}{5096} & \multicolumn{2}{l|}{23503936} & \multicolumn{2}{l|}{41227936} &  \multicolumn{2}{l|}{313253568} & \multicolumn{2}{l|}{815215040}  & \multicolumn{2}{l|}{812711}  & \multicolumn{2}{l|}{1099440} \\ 
1 & 2460  &  $(48\%)$ & 8631124  &  $(37\%)$ & 11964605  &  $(29\%)$ & 116548496 &  $(37\%)$ & 254604944  &  $(31\%)$ & 363997 &  $(45\%)$ & 409825  &  $(37\%)$\\ 
2  & 1835  &  $(75\%)$ & 4744277  &  $(55\%)$ & 622729  &  $(5\%)$ & 80952640 & $(69\%)$ & 19777422 & $(8\%)$ & 238115 & $(65\%)$ & 28732 & $(7\%)$\\ 
3  & 1768 & $(96\%)$ & 4649472 & $(98\%)$ & 608376 & $(98\%)$ & 79378192 & $(98\%)$ & 19324588 & $(98\%)$ & 232430 & $(98\%)$ & 27220 & $(95\%)$\\ 
4   & 438 & $(25\%)$ & 499215 & $(11\%)$ & 48787 & $(8\%)$ & 76163 & $(0\%)$ & 10310 & $(0\%)$ & 7405 & $(3\%)$ & 608 & $(2\%)$\\ 
5 & 336 & $(77\%)$ & 230068 & $(46\%)$ & 9412 & $(19\%)$ & 30275 & $(40\%)$ & 2181 & $(21\%)$ & 4035 & $(54\%)$ & 149 & $(24\%)$\\ 
6 & 221 & $(66\%)$ & 66128 & $(29\%)$ & 1182 & $(13\%)$ & 7766 & $(26\%)$ & 649 & $(30\%)$ & 901 & $(22\%)$ & 13 & $(8\%)$\\ 
7 & 219 & $(99\%)$ & 64744 & $(98\%)$ & 1182 &  & 7707 & $(99\%)$ & 648 &  & 815 & $(90\%)$ & 13 & \\ 
8 & 215 & $(98\%)$ & 60848 & $(94\%)$ & 1115 & $(94\%)$ & 7364 & $(96\%)$ & 583 & $(90\%)$ & 788 & $(97\%)$ & 12  & \\ 
9  & 179 & $(83\%)$ & 36085 & $(59\%)$ & 534 & $(48\%)$ & 3390 & $(46\%)$ & 253 & $(43\%)$ & 667 & $(85\%)$ & 9 & $(68\%)$\\ 
10 & 175 & $(98\%)$ & 34991 & $(97\%)$ & 533  &  & 3066 & $(90\%)$ & 253 &  & 659 & $(99\%)$ & 8 & $(97\%)$\\ 
11 & 129 & $(74\%)$ & 13131 & $(38\%)$ & 123 & $(23\%)$ & 1465 & $(48\%)$ & 65 & $(26\%)$ & 123 & $(19\%)$ & 1 & $(15\%)$\\ 
12 & 110 & $(86\%)$ & 4202 & $(32\%)$ & 26 & $(21\%)$ & 369 & $(25\%)$ & 11 & $(17\%)$ & 72 & $(58\%)$ & 1 & $(95\%)$\\ 
13  & 68 & $(62\%)$ & 1182 & $(28\%)$ & 13 & $(50\%)$ & 110 & $(30\%)$ & 6 & $(49\%)$ & 25 & $(35\%)$ & 1 & $(64\%)$\\ 
\hline 
\end{tabular}
}
\caption{\label{tresults} 
Number of expected events for 100 fb$^{-1}$ of integrated luminosity for a 115~GeV/c$^{2}$ mass Higgs and background samples at each
step of the selection process. The Cut \# is defined in Table \ref{T:cutlist}.} 
\end{center} 
\end{table*}

\begin{table}[h]
\begin{center}
\begin{tabular}{|c|c|c|c|}\hline
Estimated & \multicolumn{3}{c|}{Higgs Mass}\\ 
Results & 115 	& 125 & 135 \\  \hline \hline
Signal Events ($S$)		& 68$\pm$1 	& 58$\pm$1 	& 33$\pm$ 1 \\
Background Events ($B$)	 	& 1337$\pm$35 & 1341$\pm$38 	& 1210$\pm$37 \\
$S/\sqrt{B}$	 	& 1.86$\pm$0.06 & 1.58$\pm$0.05 & 0.95$\pm$0.04 \\
\hline \hline
\end{tabular}
\caption{\label{t:vbfph_sig_mass} Expected signal and background events and overall significance uncertainties for Higgs mass of 115, 125, and 135~GeV/$c^2$ for 14 TeV collision energy.}
\end{center}
\end{table}

\begin{table}[htdp] 
\begin{center} 
\begin{tabular}{|l|c|}\hline 
Cut \# & Selection Criteria \\ \hline
1 & $p_{\rm{T}}(\gamma$) $> 30$ GeV/$c$ \\ 
2 & \# of Jets $\geq 4$ \\
3 & \# of Central Jets $\geq 2$ \\
4 & Two b-jets  \\
5 & $p_{\rm{T}}(j1)>$55 GeV/$c$ \\ 
6 & M$(j1,j2)>$695.0 GeV/c$^2$ \\
7 & $\Delta \eta(j1,j2)>3.25$ \\
8 & $\theta(b1,b2)<0.92$\\
9 & $\Delta \eta(b1,b2)<1.25$\\
10 & $\eta(b1) \times \eta(b2)>-0.25$\\ 
11 &  M$(b1,b2)>$100 GeV/c$^{2}$\\
12 & M$(b1,b2)<$125 GeV/c$^{2}$\\
13 & Central Jet Veto   \\ \hline
\end{tabular}
\caption{\label{T:cutlist} Selection Criteria}
\end{center}
\end{table}

\section*{References}
\bibliography{refer}

\end{document}